\documentclass[prc,aps,twocolumn,showpacs,letterpaper]{revtex4}

\begin{document}

\title{Fragment Isospin as a Probe of Heavy-Ion Collisions}

\author{H. Xu}
\author{R. Alfaro}
\author{B. Davin}
\author{L. Beaulieu}\altaffiliation{Present address: Universite Laval, Quebec, Canada.}
\author{Y. Larochelle}\altaffiliation{Present address: Universite Laval, Quebec, Canada.}
\author{T. Lefort}\altaffiliation{Present address: Universite de Caen, Caen, France.}
\author{R. Yanez}\altaffiliation{Present address: Universidad de Chile, Santiago, Chile.}
\author{R.T. de Souza}
\affiliation{
Department of Chemistry and Indiana University Cyclotron Facility, \\ 
Indiana University, Bloomington, IN 47405} 

\author{T.X. Liu}
\author{X.D. Liu}
\author{W.G. Lynch}
\author{R. Shomin}
\author{W.P. Tan}
\author{M.B. Tsang}
\author{A. Vander Molen}
\author{A. Wagner}\altaffiliation{Present address: Institute of Nuclear and Hadron Physics,
Forschungszentrum, Rossendorf, Dresden, Germany. }
\author{H.F. Xi}
\author{C.K. Gelbke}
\affiliation{
National Superconducting Cyclotron Laboratory and Department of
Physics and Astronomy,
Michigan State University, East Lansing, MI 48824}

\author{R.J. Charity}
\author{L.G. Sobotka}
\affiliation{
Department of Chemistry, Washington University, St. Louis, MO 63130}

\author{A.S. Botvina}
\affiliation{Gesellschaft f{\"u}r Schwerionenforschung, 64291 Darmstadt, 
Germany and \\
Institute for Nuclear Research, Russian Academy of Science, 117312 Moscow, 
Russia}

\date{\today}

\begin{abstract}
Isotope ratios of fragments produced at mid-rapidity in peripheral and
central collisions of $^{114}$Cd ions with $^{92}$Mo and $^{98}$Mo  
target nuclei at E/A = 50 MeV are compared. Neutron-rich isotopes are 
preferentially produced in central collisions as compared to  
peripheral collisions.
The influence 
of the size (A), density, N/Z, E$^*$/A, and E$_{flow}$/A 
of the emitting source on the measured isotope ratios was explored by 
comparison with a statistical model (SMM). 
The mid-rapidity region associated with peripheral collisions does
{\em not} appear to be neutron-enriched relative to central collisions.

\end{abstract}


\maketitle

When projectile and target nuclei collide at intermediate 
energies (20$\leq$E/A$\leq$100 MeV), a
characteristic signature is the copious  
production of intermediate mass fragments (IMF:3$\leq$Z$\leq$20). 
Continuum models of nuclear matter have recently been used to propose that
nuclear matter at high excitation may fractionate\cite{Muller95}, 
that is separate into liquid and gaseous phases with different N/Z. 
Driven by the density 
dependence of the asymmetry energy, such a process predicts formation 
of a neutron-rich gas and a proton-rich liquid\cite{xu1}.
For peripheral and mid-central collisions, where a low 
density zone between the projectile and target nuclei is formed,
the density 
dependence of the asymmetry energy may also result in neutron enrichment of 
the mid-rapidity region. 
Dynamical stochastic transport models make specific predictions
for neutron enrichment 
with increasing centrality\cite{Sobotka94,diToro00}. 
It is therefore interesting to 
examine the isotopic composition of fragments at mid-rapidity and determine 
the dependence of
neutron enrichment on centrality. 
Although the elemental breakup of the mid-rapidity
zone in peripheral collisions has been previously 
investigated\cite{Montoya94}, only recently has the N/Z  
composition of IMFs in this region become available.   
While investigation of similar
phenomena at lower incident energies \cite{Casini93,Toke95} suggests that  
fragment emission might have a non-equilibrium character,
it is instructive to first examine the experimental
findings within a statistical context. 

For peripheral and mid-central collisions of two heavy-ions at intermediate 
energies, it is well established that the 
nature of the collision is dissipative and largely 
binary\cite{Lott92,Beaulieu96,Plagnol00}. 
Following the interaction phase in which the 
kinetic energy of relative motion is converted into intrinsic excitation 
(manifested by the velocity damping of the projectile) and 
pre-equilibrium emission, the excited  
projectile-like (PLF) and target-like (TLF)
reaction partners decay by statistical emission of 
neutrons and light charged particles. 
When the IMF yield is examined, however, one observes a large 
excess over the sequential decay component located
at mid-rapidity between the PLF and 
TLF\cite{Montoya94,Toke95,Lukasik97,Plagnol99}. 
For light clusters (Z=1,2), 
neutron-rich clusters are particularly prevalent at 
mid-rapidity\cite{Dempsey96}. 
Whether this neutron enrichment of light clusters signals an 
isospin fractionation of the system remains a topic of much debate.

    In order to probe the N/Z of the mid-rapidity region, we investigated 
IMF and light charged particle (LCP:1$\leq$Z$\leq$2) emission in the reaction
$^{114}$Cd + $^{92}$Mo and  $^{114}$Cd + $^{98}$Mo at E/A=50 MeV. In an 
exclusive 4$\pi$ setup we detected isotopically identified LCPs and IMFs 
with Z$\leq$9 in the angular range 
7$^{\circ}$$\leq$$\theta_{lab}$$\leq$58$^{\circ}$. These
fragments were detected with the high resolution silicon-strip array LASSA
(Si(IP)-Si(IP)-CsI(Tl)/PD)
which had an energy threshold of 2 and 4 MeV/u for $\alpha$ and carbon 
fragments, respectively\cite{davin01,Wagner01}. Forward-moving projectile-like fragments were 
identified in an annular Si(IP)/CsI(Tl) ring detector 
2.1$^{\circ}$$\leq$$\theta_{lab}$$\leq$4.2$^{\circ}$,
which provided unit Z resolution for Z$\leq$48. Additionally, charged particles
were measured in the range 5$^{\circ}$$\leq$$\theta_{lab}$$\leq$168$^{\circ}$
by the Miniball/Miniwall array, allowing global event characterization
\cite{Miniball}.

We examined the isotopic composition of fragments at mid-rapidity  
by first selecting collisions on the basis of the charged particle 
multiplicity, N$_C$.
When N$_C$$\leq$13 (``peripheral collisions'';$\langle$N$_C$$\rangle$=9.9),
a fragment emission pattern is observed that is consistent with fragment 
emission following
a dissipative, largely binary collision between the projectile and target 
nuclei. 
In contrast, when N$_C$$>$19 
(``central'' collisions;$\langle$N$_C$$\rangle$=22.2), 
the emission pattern 
for He and Li fragments is broad and featureless with
substantial emission near the center-of-mass velocity.
By relating the multiplicity distribution
to a reduced impact parameter scale\cite{Cavata}, we deduced that the 
multiplicity interval N$_C$ $\le$13 corresponds to 
$\langle$b/b$_{max}$$\rangle$ = 0.65, while 
N$_C$ $\ge$20 corresponds to $\langle$b/b$_{max}$$\rangle$ = 0.26.

\begin{figure} \vspace*{3.4in}
\includegraphics{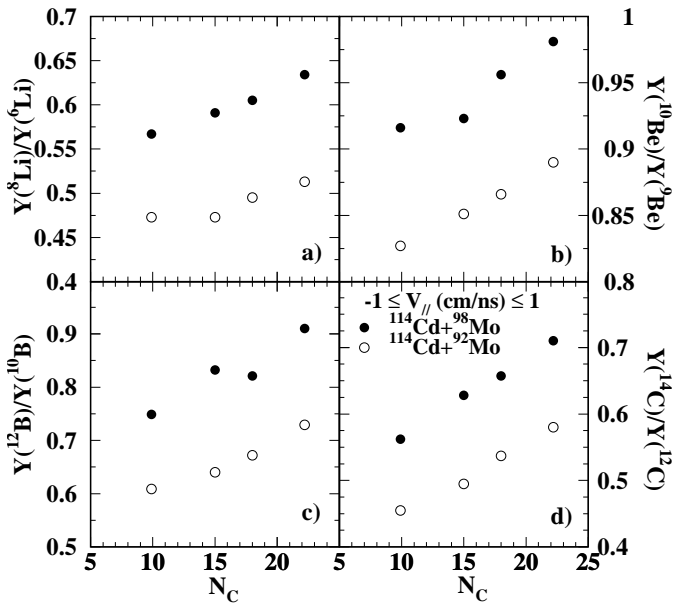}
\caption[]
{Ratio of isotopic yields for fragments with 3$\leq$Z$\leq$6 at 
mid-rapidity (-1$\leq$v$_{//}$(cm/ns)$\leq$1) as a function of 
N$_c$ for the reactions $^{114}$Cd + $^{92}$Mo and 
$^{114}$Cd + $^{98}$Mo at E/A = 50 MeV. 
} 
\end{figure}

We further selected fragments by focusing on those 
detected in the interval 
-1$\leq$V$_{//}$(cm/ns)$\leq$1, where 
V$_{//}$ is the longitudinal component 
of a fragment's velocity in the center-of-mass, and examined different 
isotope ratios as a function of N$_C$. 
We subsequently utilize the 
nomenclature
HH (heavy-heavy) and HL (heavy-light) to represent 
the  $^{114}$Cd + $^{98}$Mo  and 
$^{114}$Cd + $^{92}$Mo reactions, respectively. 
It should be noted 
that for the HH system, the N/Z of the projectile and target is 
essentially the same ((N/Z)$_{proj}$=1.375;(N/Z)$_{targ}$=1.333) thus 
providing little driving force for neutron enrichment of mid-rapidity due to 
isospin equilibration.

The systematic isotopic behavior at mid-rapidity is 
shown in Fig.~1 where we explore the
the dependence of mid-rapidity neutron enrichment on both  
centrality and the neutron content of the original system by examining  
isotope ratios for 3$\leq$Z$\leq$6. Solid symbols 
depict the dependence of the yield ratio Y($^{A2}$Z)/Y($^{A1}$Z), 
where A2$>$A1,
for the neutron-rich HH system  while open symbols 
represent the same ratio for the relatively neutron-deficient HL
system. 
In all cases shown in Fig.~1, a 
roughly linear increase of the relative yield for neutron-rich fragments
as a function of 
on N$_C$ is observed.  
As large N$_C$ can be associated with more central,
higher excitation collisions, it appears that 
neutron-rich fragments at mid-rapidity are 
preferentially produced under these conditions.
If the N/Z at mid-rapidity for peripheral and central collisions is the same,
and secondary decay contributions are comparable for the two cases, 
one would expect the relative probability for neutron-rich fragments as 
compared to neutron-poor fragments to be constant. Should secondary decay 
increase in importance for more central collisions due to higher excitation,
neutron-rich fragments would be favored in peripheral collisions as compared 
to central collisions.
This observation
that neutron-rich fragments are preferentially produced in central collisions
might suggest that peripheral collisions are characterized by a lower N/Z than 
central collisions.
However, it is also conceiveable that higher excitation for peripheral 
collisions supresses the survival of neutron-rich fragments. 

It is evident in Fig.~1 that 
while both the HH and HL systems manifest the same trend with N$_C$, the 
relative yield of the neutron-rich species is enhanced in the HH system as 
compared to the HL system. Moreover, the enhancement observed for the HH 
system as compared to the HL system for large N$_C$ that has been previously 
observed\cite{xu1,Kunde96} occurs with 
essentially the same magnitude even for the peripheral (low N$_C$) case.
This result indicates that the initial difference in N/Z between the HH 
and HL systems influences the relative production of fragments 
at mid-rapidity, and persists from 
peripheral collisions to central collisions. For the following analysis 
we determined that the HH and HL systems provided essentially the same 
results. For simplicity, we therefore focus on the HH system. 

To investigate the isotopic yields associated with 
peripheral and central collisions further, we compared our results with a
statistical multifragmentation model (SMM)\cite{Bondorf95,Botvina01}. 
In this microcanonical model,
the statistical partition into fragments can be calculated for a source
characterized by a size of (A), a density ($\rho$/$\rho_0$), neutron-to-proton
ratio  (N/Z), 
with a given excitation energy (E$^*$/A), and possibly a radial flow energy
(E$_{flow}/A$). For the present analysis we have assumed 
that fragment production at mid-rapidity results from a single 
mid-rapidity 
source. This assumption of a single source, 
formed by the overlap of the projectile and target nuclei,
is somewhat simplistic. Nevertheless, it allows us to consider the general 
factors governing fragment production at mid-rapidity. It has recently 
been suggested that fragment production at mid-rapidity may be influenced,
particularly for peripheral collisions, by the Coulomb proximity-induced 
emission of
the projectile-like and target-like fragments\cite{Botvina99}. 
Detailed comparisons with such a scenario are beyond the scope of the
present investigation but are being pursued and will be presented in a
forthcoming publication. All model calculations presented have been filtered 
by the angular acceptance of LASSA 
(7$^{\circ}$$\leq$$\theta_{lab}$$\leq$58$^{\circ}$) as well as the 
restriction that -1$\leq$V$_{//}$(cm/ns)$\leq$1.

\begin{figure} \vspace*{3.4in}
\includegraphics{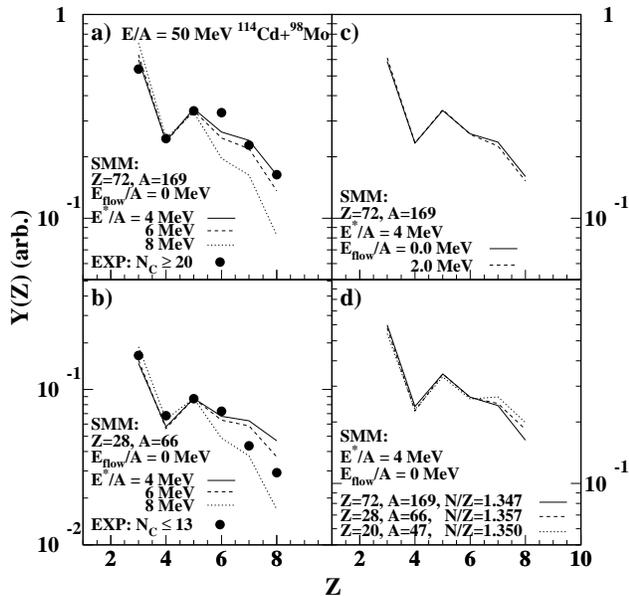}
\caption[]
{Panel a: Comparison of the measured Z distribution associated
with central collisions with the predicted Z 
distribution of the SMM model for a source with Z=72, A=169 and 
E$^*$/A = 4,6, and 8 MeV. 
Panel b: Comparison of the measured Z distribution associated
with peripheral collisions with the predicted Z 
distribution of the SMM model for a source with Z=28, A=66 and 
E$^*$/A = 4,6, and 8 MeV. 
Panel c: Investigation of the influence of radial flow on the 
Z 
distributions predicted by the SMM model for a source with Z=72, A=169, 
E$^*$= 4 MeV. E$_{flow}$/A was selected to be 0 and 2 MeV.
Panel d: Investigation of the influence of source size with approximately
constant N/Z on the  Z 
distributions as predicted by the SMM model.
} 
\end{figure}

In order to explore the sensitivity of the measured isotopic yields
to the excitation (E$^*$/A) and N/Z of the emitting source, it was
first necessary to constrain the size (A), density ($\rho$/$\rho_0$),
and radial flow energy E$_{flow}$/A in the SMM model. 
We first compared the experimentally measured 
Z distributions in the range -1$\leq$V$_{//}$(cm/ns)$\leq$1 with the 
predictions of the SMM model. The experimental Z distribution observed 
for central and peripheral
reactions are represented as the solid symbols
in Fig.~2. These data have been normalized by the number of events in 
each case. The experimental data exhibit an overall ``exponential'' decrease
in yield with increasing Z.

In the SMM calculations, for the case of central collisions,
we assumed the source 
size was 80$\%$ the size of the composite system with N/Z equal to the 
composite system. 
For this case we display the Z distribution predicted by the SMM model for 
E$^*$/A= 4,6, and 8 MeV when $\rho$/$\rho_0$=1/6. The presented 
yields have been normalized to the yield for Z=5 at 
mid-rapidity allowing comparison of the shape of the Z distribution. 
The calculations with E$^*$/A=4-6 MeV provide 
reasonable agreement with the experimental data (Fig. 2a) 
,however, for E$^*$/A=8 MeV the predicted Z distribution is significantly 
steeper than the measured one. As expected, we observe that the 
slope of the Z distribution is mainly sensitive to the excitation 
of the source.

In order to estimate the size of the mid-rapidity source associated with 
peripheral collisions, we have examined the detected charge, Z$_{sum}$, in the 
ring counter with the appropriate velocity and 
associated multiplicity N$_c$. We 
then assumed that the N/Z of this PLF was the same as the N/Z of the 
projectile. By utilizing a participant-spectator model we 
were able to estimate the contribution of
neutrons and protons to mid-rapidity from the target nucleus.  
By this means, we estimated the size of the mid-rapidity source to be A=66.

For the case of 
peripheral collisions (Fig. 2b), with $\rho$/$\rho_0$=1/6 and
a source size of A=66, Z=28 (N/Z=1.357), we show
the extent to which the Z 
distribution is modified as E$^*$/A increases. 
For this source an excitation energy E$^*$/A $\sim$ 6-8 MeV 
provides the best agreement with the experimental results. 

We have also explored the sensitivity of the Z 
distribution to the source N/Z and its size. While selection of a 
smaller N/Z (neutron-poor) source does supress the yield for larger Z 
fragments, the overall modification of the Z distribution for the elements 
shown is not strong. The influence of radial flow, 
reported in similar reactions, on our results has also been examined. 
As shown in Fig.~2c,
the Z distribution predicted by the SMM is 
essentially insensitive to this collective motion i.e. the fragment velocities
are already well above the experimental acceptance.
The influence of source size (assuming a near constant N/Z) is shown in 
Fig.~2d. For the range of source sizes assumed, which span the
physically realistic range, the Z distribution is fairly independent of the 
source size with only a minor sensitivity for the largest fragments.

\begin{figure} \vspace*{3.4in}
\includegraphics{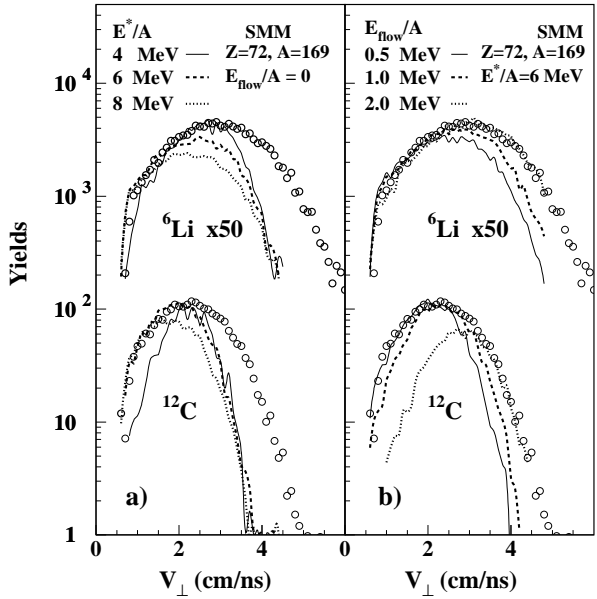}
\caption[]
{Panel a: Comparison of V$_{\perp}$ distributions 
associated with central collisions for $^6$Li and $^{12}$C fragments with 
the SMM predictions for a source (Z=72,A=169,E$_{flow}$/A=0 MeV)
with E$^*$/A=4,6, and 8 MeV.
Panel b: Comparison of V$_{\perp}$ distributions 
associated with central collisions for $^6$Li and $^{12}$C fragments with 
the SMM predictions for a source (Z=72,A=169,E$^{*}$/A=6 MeV)
with different amounts of radial flow energy, E$^*$/A=0.5,1.0, and 2 MeV.
} 
\end{figure}

We also investigated whether the transverse velocity distributions,V$_\perp$, 
for fragments emitted at mid-rapidity could be used to constrain the
E$^*$/A of the emitting source and determine if any radial flow is present  
for this reaction.
Distributions for $^6$Li and 
$^{12}$C, representative of other fragments, are shown in Fig.~3. 
In panel a), the experimental V$_\perp$ distributions 
of $^6$Li and 
$^{12}$C fragments measured for 
central collisions (N$_C$$\ge$20) are compared with the SMM 
calculations for the decay of
the large source (Z=72, A=169) with E$_{flow}$/A = 0 for E$^*$/A= 4,6, 
and 8 MeV. 
Most clearly in the case of 
$^{12}$C, increased excitation energy results in a better reproduction
of the low velocity 
portion of the V$_\perp$ distributions. 
While E$^*$/A = 4 MeV appears to be too low an excitation
to properly reproduce the low velocity portion of the spectrum,  
E$^*$/A = 6-8 MeV provides an adequate description. In  
Fig.~3b, we demonstrate the influence of E$_{flow}$/A on the
V$_\perp$ distribution. For the case of $^6$Li, 
increasing E$_{flow}$/A from 
0.5 to 2 MeV provides a better description of the high velocity tail of the
V$_\perp$ distribution. The dramatic influence of E$_{flow}$/A for 
heavy fragments is evident for $^{12}$C where one observes that inclusion of 
a large flow for such fragments substantially shifts the V$_\perp$ 
distribution to higher values of V$_\perp$, therefore resulting 
in a significant 
underprediction on the low V$_\perp$ side. 
The difficulty in describing the entire
V$_\perp$ distribution for several different elements with a 
single value for the
radial expansion energy is hardly surprising 
considering that in reality we deal with an ensemble of sources. 
In reality, all quantities 
such as
E$_{flow}/A$ are probably distributions which may in fact depend on 
particle type. In this analysis we only attempt to describe the 
average behavior
which affects the bulk of the cross-section. As indicated in 
Fig.~3, the bulk of the cross-section in  
the V$_\perp$ distributions for $^6$Li and $^{12}$C are consistent with 
zero to modest flow energy i.e. E$_{flow}/A$ = 0-0.5 MeV.

Having constrained the source size, the excitation energy, and the flow 
energy for both the peripheral and 
central cases at mid-rapidity, we examined the sensitivity of the
ratio Y$_{cntr}$/Y$_{peri}$ 
as a function of fragment neutron number N, for 
fragments with Z=3-6. The experimental results and the corresponding
SMM calculations are presented in Fig.~4. 
The measured and predicted yields of each element for
both peripheral and central collisions have  been individually normalized, 
thus providing a comparison of isotopic predictions that is insensitive to the 
prediction of elemental yields.
One observes that in all cases, the quantity Y$_{cntr}$/Y$_{peri}$ 
for the experimental data manifests a 
general increase with increasing neutron number.

\begin{figure} \vspace*{3.6in}
\includegraphics{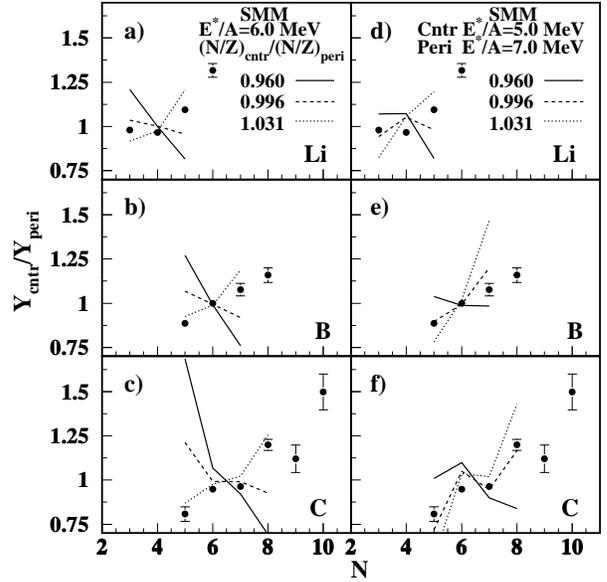}
\caption[]
{Relative yield at mid-rapidity for isotopes of 
Li, B, and C associated with central and peripheral collisions. 
Panels a-c): SMM calculations assume both peripheral and central sources have
E$^*$/A = 6 MeV and E$_{flow}$/A = 0. 
Panels d-f): The central and peripheral sources are assumed to have E$^*$/A=5 and 7 MeV, respectively, and E$_{flow}$/A = 0. 
Solid, dashed, and dotted lines 
represent varying degrees of neutron enrichment of the peripheral source as 
compared to the central source. 
} 
\end{figure}

Following our constraint of the E$^*$/A and E$_{flow}$/A
as described above, we have chosen two different frameworks within which 
to examine our results. In the first case, depicted in Fig.~4a-c,  
we assumed that the E$^*$/A of 
the mid-rapidity region for both peripheral and central collisions is the 
same, i.e. E$^*$/A = 6.0 MeV, with a zero radial flow. 
The Z  and A of the central source 
was fixed at Z=72 and A=169. For the peripheral source, the size of the source
was taken to be A=66 while the Z was varied between Z=27-29. 
The solid, dashed, and dotted lines indicate the results when 
(N/Z)$_{cntr}$/(N/Z)$_{peri}$ = 0.960,0.996, and 1.031, respectively, i.e. the 
peripheral source changes from being neutron-rich to being neutron-deficient 
as compared to the central source. 
The overall behavior of the SMM calculations can be summarized as follows:
The slope of the quantity Y$_{cntr}$/Y$_{peri}$ as a function of neutron number
depends on the ratio of the N/Z of the central and peripheral sources. As
one goes from a neutron-rich peripheral source to a neutron-deficient 
peripheral source (with a constant N/Z central source), the slope changes 
from negative to positive.
The experimental data indicates that within this framework the peripheral 
source is slightly neutron-deficient as compared to the central source.

In Fig.~4d-f we present a second scenario in which we allow the E$^*$/A of the
peripheral and central cases to be different. For this scenario, the  
E$^*$/A was determined by comparing the predicted Z distributions to 
the experimental data. As a result of these comparisons, 
we deduced E$^*$/A = 5.0 MeV for the central case 
and E$^*$/A = 7.0 MeV for the 
peripheral case. Under the assumption of these two different temperatures,
comparison with the SMM calculations indicates that overall 
the data is compatible only with the case of no significant 
neutron enrichment of
the peripheral source in comparison to the central source. 
In both scenarios presented we have determined 
that our conclusions are independent 
of our assumptions of the breakup density, source size, and radial flow.

Within a strictly geometrical picture of the collision the N/Z of the 
system remains constant in the overlap region. 
As indicated by the comparison of the measured Z distributions 
with the SMM calculations, the initial excitation of both the peripheral and 
central sources may well be comparable. 
Secondary decay effects are thus not expected 
to dominate the behavior of the ratio Y$_{central}$/Y$_{peripheral}$ 
with neutron number. 
Rather, the trend of this ratio with neutron number
may suggest that the mid-rapidity source that produces fragments in peripheral
collisions is 
neutron-deficient as compared to the mid-rapidity source for central 
collisions. Alternatively, the mid-rapidity source in peripheral and 
central collisions may have the same N/Z. In this case, however, the E$^*$/A
of the peripheral source must exceed the E$^*$/A of the central source. 

While at the present time neither possibility can be eliminated, it
is interesting to speculate on the possible reason the mid-rapidity
region may be neutron-deficient for peripheral collisions as compared to 
central collisions {\em as evinced by fragment isotope yields}.
One scenario consistent with these results is that the initial source 
at mid-rapidity
for both the peripheral and central cases is consistent with the N/Z of the 
system. Prior to fragment formation, however, neutrons are preferentially 
depleted from this system thus affecting the N/Z ratio. As shown in Fig.~4, 
removal of a single neutron has a measurable impact on the observed 
isotopic yields.

In summary, we have probed the N/Z composition at mid-rapidity for both
peripheral and central collisions by examining the N/Z ratio of fragments
with 3$\leq$Z$\leq$6. 
Neutron-rich fragments are preferentially produced for central collisions as
opposed to peripheral collisions. 
This supression of neutron-rich fragments
in peripheral collisions can be related to either a neutron-deficient source 
at mid-rapidity  for
peripheral collisions as compared to central collisions 
or a more excited mid-rapidity region in peripheral collisions.

	We would like to acknowledge the 
valuable assistance of the K1200 cyclotron staff at MSU-NSCL for
providing the high quality beams that made this experiment possible. We are\
grateful to Dr. Jan Toke for his help with the data acquisition. 
One of the authors (RdS) is especially grateful to 
Commissariat a l' Energie Atomique and G.A.N.I.L. (France) 
for support enabling this work during a sabbatical leave. A.S.B. thanks 
Indiana University for warm hospitality and support during his visit.
This work was supported by the
U.S. Department of Energy under DE-FG02-92ER40714 (IU), 
DE-FG02-87ER-40316 (WU) and the
National Science Foundation under Grant No. PHY-95-28844 (MSU).\par



\end{document}